\documentclass[10pt,conference]{IEEEtran}
\IEEEoverridecommandlockouts
\usepackage{cite}
\usepackage{amsmath,amssymb,amsfonts}
\usepackage{algorithm}
\usepackage{graphicx}
\usepackage{textcomp}
\usepackage{xcolor}
\usepackage{algpseudocode}
\usepackage{hyperref}

\newdimen{\algindent}
\newcommand{\commentsymbol}{//}
\algrenewcommand\algorithmiccomment[1]{\hfill \commentsymbol{} #1}
\newdimen{\algindent}
\hypersetup{hidelinks,linkcolor = black} % Changes the link color to black and hides the hideous red border that usually is created

\def\BibTeX{{\rm B\kern-.05em{\sc i\kern-.025em b}\kern-.08em
    T\kern-.1667em\lower.7ex\hbox{E}\kern-.125emX}}
\begin{document}

\newcommand{\QD}{QuickDraw}
\newcommand{\dataName}{DoodleUINet}
\newcommand{\toolName}{PSDoodle}
\newcommand{\appName}{Doodle2App}

\newcommand{\toolwithtext}{TpD}

\title{Searching Mobile App Screens via Text + Doodle}

\author{\IEEEauthorblockN{1\textsuperscript{st} Soumik Mohian}
\IEEEauthorblockA{\textit{dept. computer science and engineering} \\
\textit{University of Texas at Arlington}\\
Arlington, USA \\
soumik.mohian@mavs.uta.edu}
\and
\IEEEauthorblockN{2\textsuperscript{nd} Christoph Csallner}
\IEEEauthorblockA{\textit{dept. computer science and engineering} \\
\textit{University of Texas at Arlington}\\
Arlington, USA \\
csallner@uta.edu}

}

\maketitle

\begin{abstract}

Locating a specific mobile application screen from existing repositories is restricted to basic keyword searches, such as Google Image Search, or necessitates a complete query screen image, as in the case of Swire. However, interactive partial sketch-based solutions (PSDoodle) have limitations, including inaccuracy and an inability to consider text appearing on the screen. A potentially effective solution involves implementing a system that provides interactive partial sketching and should further incorporate text queries to enhance its capabilities. Our approach, \toolwithtext{}, represents the pioneering effort to enable an iterative search of screens by combining interactive sketching and keyword search techniques. \toolwithtext{} is built on a combination of the Rico repository of approximately 58k Android app screens and the \toolName{}. Our evaluation with third-party software developers revealed that \toolwithtext{} demonstrated equivalent top-10 screen retrieval accuracy compared to the state-of-the-art Swire. Additionally, it provided higher accuracy and reduced screen retrieval time compared to the interactive sketch-based query solutions provided by \toolName{}.

\end{abstract}


\section{Introduction}

Traditionally, When people want to find mobile app examples, they usually use keywords to search for it on tools like Google's image search\cite{bernal2019guigle,kolthoff2023data}. However, it can be tough to describe the structure of a screen accurately with just words. Sketch-based search presents a structured approach for representing element structure with greater ease. Through this approach, users can search mobile screens by creating visual representations of either the entire screen or a portion~\cite{mohian2022psdoodle, huang2019swire, sain2020cross}.

A complete screen drawing-based solution can be tedious, slow, and lack interactivity. Partial sketching as an interactive alternative for search has drawbacks in terms of precision and can yield inconsistent outcomes depending on the user's proficiency. ~\cite {mohian2022psdoodle}. The sketch-based search method has yet to be widely embraced and still encounters obstacles in identifying textual elements on the screen or differentiating between two text buttons with varying content. No attempt was made to merge these two distinct search approaches.

%% Why is it interesting \& important?
As mobile applications gain widespread use, companies allocate significant resources towards developing user interfaces that align with user experience standards~\cite{ines2017evalmobileinterface,hellmann2011ruletestUI, user_interface_growth_2028}. A search engine that efficiently and accurately locates mobile app screens can offer substantial benefits for user interface design, such as facilitating requirement gathering, trend analysis, feature analysis, developer creativity, and evaluation benchmarks. Consequently,  an effective mobile app screen search engine can significantly impact software developers and end-users.

%% Solution

\toolwithtext{} represents a novel approach that combines sketch and text to search for mobile screens. \toolwithtext{} features a digital drawing interface similar to \toolName{}, which allows users to draw using a touch or a mouse. Moreover, \toolwithtext{} provides keyword-based searching for mobile screens. By integrating word and doodle queries, \toolwithtext{} retrieves more relevant mobile app screens from a pool of 58k Rico screens~\cite{deka2017rico}.

\begin{figure*}[h!t]
 \centering
 \includegraphics[width=\linewidth] {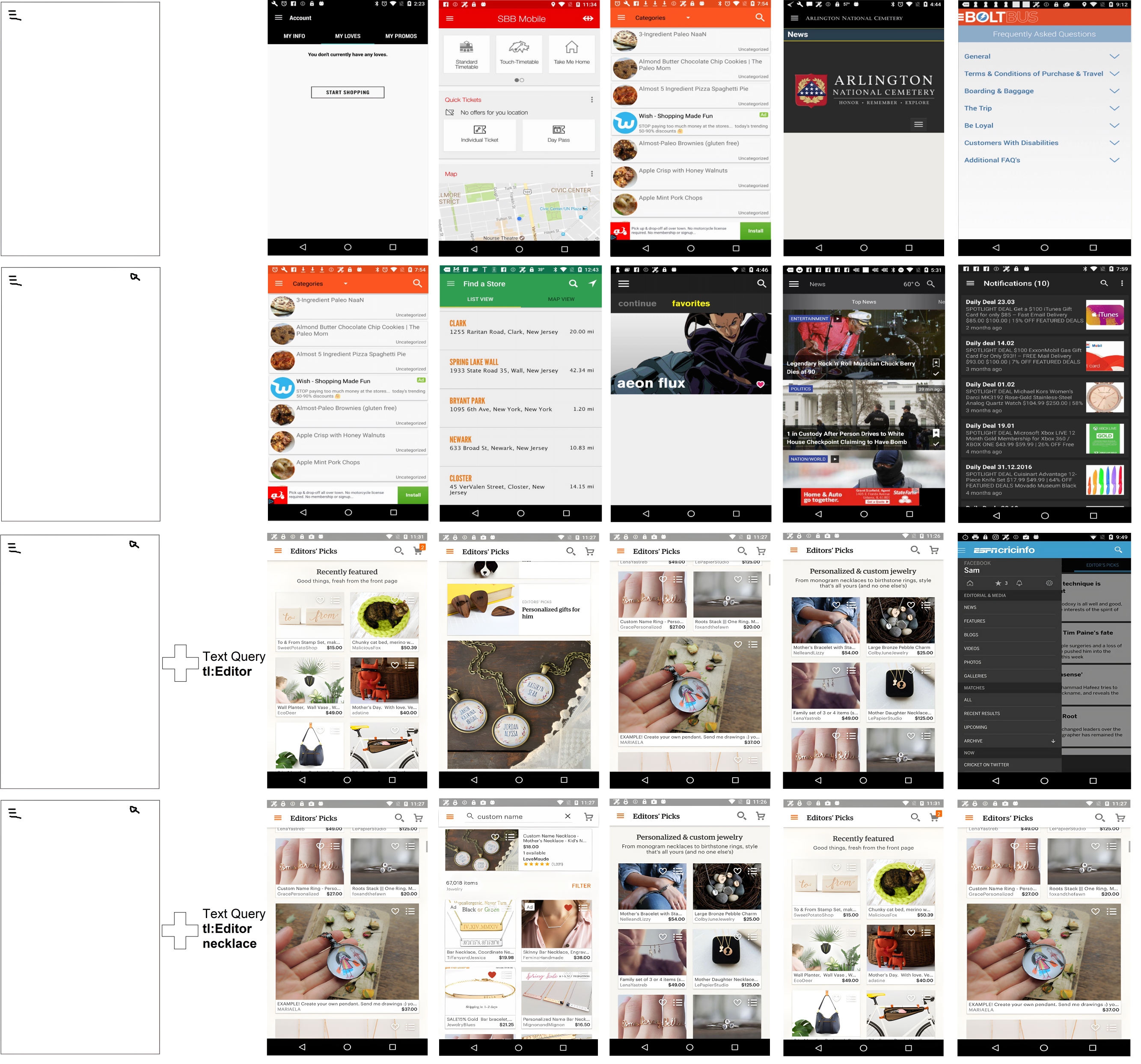}
 \caption{A screen query session with \toolwithtext{} and the first column of each row displays the search. The subsequent columns in each row show the top-5 search results (in order) out of 58k Rico screens returned by \toolwithtext{}. All screens exhibit the sketched UI elements in roughly the exact location as the user drew them. When the user adds text to the query, the updated search results show screens containing the sketched UI elements and the specified text in their respective locations. Within each row, \toolwithtext{} returns the top 50 ranked result screens in just 2 seconds, including the round-trip duration from the user's device to the AWS-hosted \toolwithtext{}.  
 }
 \label{fig:iterativeRetriveval_Text} 
\end{figure*}

The image in Figure~\ref{fig:iterativeRetriveval_Text} depicts an example query session of \toolwithtext{}. The user initially draws a 'menu' icon on the top left corner of the canvas in the \toolwithtext{} website. After completing the drawing of the 'menu' icon, \toolwithtext{} retrieves the top 50 best matching mobile screens by iterating through 58k Rico screens and displays them at the bottom of the canvas within 2 seconds. The first row of the image in Figure~\ref{fig:iterativeRetriveval_Text} highlights the top 5 screens of \toolwithtext{}, and all of them have a 'menu' icon situated near the area where the user sketched it. Next, the user adds a 'search' icon doodle in the top right corner of the canvas, and \toolwithtext{} promptly updates the search result within 2 seconds. The second row of the image in Figure~\ref{fig:iterativeRetriveval_Text} displays the updated top 5 screens, and all contain both the 'menu' and 'search' icons in their respective positions. Afterward, the user includes a text query using the term 'Editor' and marks its position as top-left on the screen with the syntax 'tl: Editor.' The third row of the image in Figure~\ref{fig:iterativeRetriveval_Text} showcases the top 5 results retrieved by \toolwithtext{}, all of which feature the text 'Editor' in the top-left corner, along with the 'menu' and 'search' icons situated in the positions specified by the sketch query. Finally, the user adds another text query, 'necklace,' prompting \toolwithtext{} to update the search results. The last row of the image in Figure~\ref{fig:iterativeRetriveval_Text} displays the top 5 search results.

\toolName{}'s query can be keywords, sketch, or both. \toolwithtext{} incorporates \toolName{}'s deep learning techniques to recognize sketched UI elements on the drawing interface. \toolwithtext{} also adopts \toolName{}'s algorithm to compute a ranking score for a Rico screen based on UI element type, position, and shape. For keyword query, \toolwithtext{} matches terms with the mobile screen's visible text and UI element descriptions while accounting for synonymous and semantically related words. The tool is invariant to inflectional or derivational-related changes in the text. \toolwithtext{} aware of the position of the text on the mobile screen, allowing users to annotate a text query with the 'position: keyword' syntax. The position can take one of the 12 values (l: left, r: right, t: top, b: bottom, lt/tl: top-left, rt/tr: top-right, bl/lb: bottom-left, br/rb: bottom-right). \toolwithtext{} fetches search results in real-time for any update in query, providing an interactive and iterative approach to screen search.

We asked 10 participants who were new to the \toolwithtext{} to search for a specific Rico mobile screen using text and sketches. \toolwithtext{} successfully retrieved and displayed the target screen in its top-10 search results 90\% of the time, with the target screen consistently appearing on the first page (each page exhibits 50 mobile screens ) at the end of the query. The average query length was 4, and the average search time was 45 seconds. These outcomes showcase significant progress in performance compared to the closest related tool, \toolName{}~\cite{mohian2022psdoodle,mohian2022psdoodle_demo}. Moreover, the achieved accuracy levels are on par with those reported in work by Swire\cite{huang2019swire}. This paper offers several significant contributions-

\begin{itemize}
\item  \toolwithtext{} is the first tool, that interactive and iterative search system for mobile screens that incorporates both text and sketch input. Users can access the tool at the following URL: \url{http://pixeltoapp.com/WnD}.

\item In comparison to the state-of-the-art, \toolwithtext{} exhibits equivalent top-10 search accuracy while significantly outperforming in terms of speed.

\item The source code, processing scripts, training data, and experimental results of \toolwithtext{} are accessible under open-source licenses.~\footnote{https://github.com/DoodleUI/WnD}

\end{itemize}

\section{Background}

\subsection{Rico: Corpus of 72k Android App Screens}

\begin{figure*}[h!]
\centering
\includegraphics[width=.8\linewidth,clip] {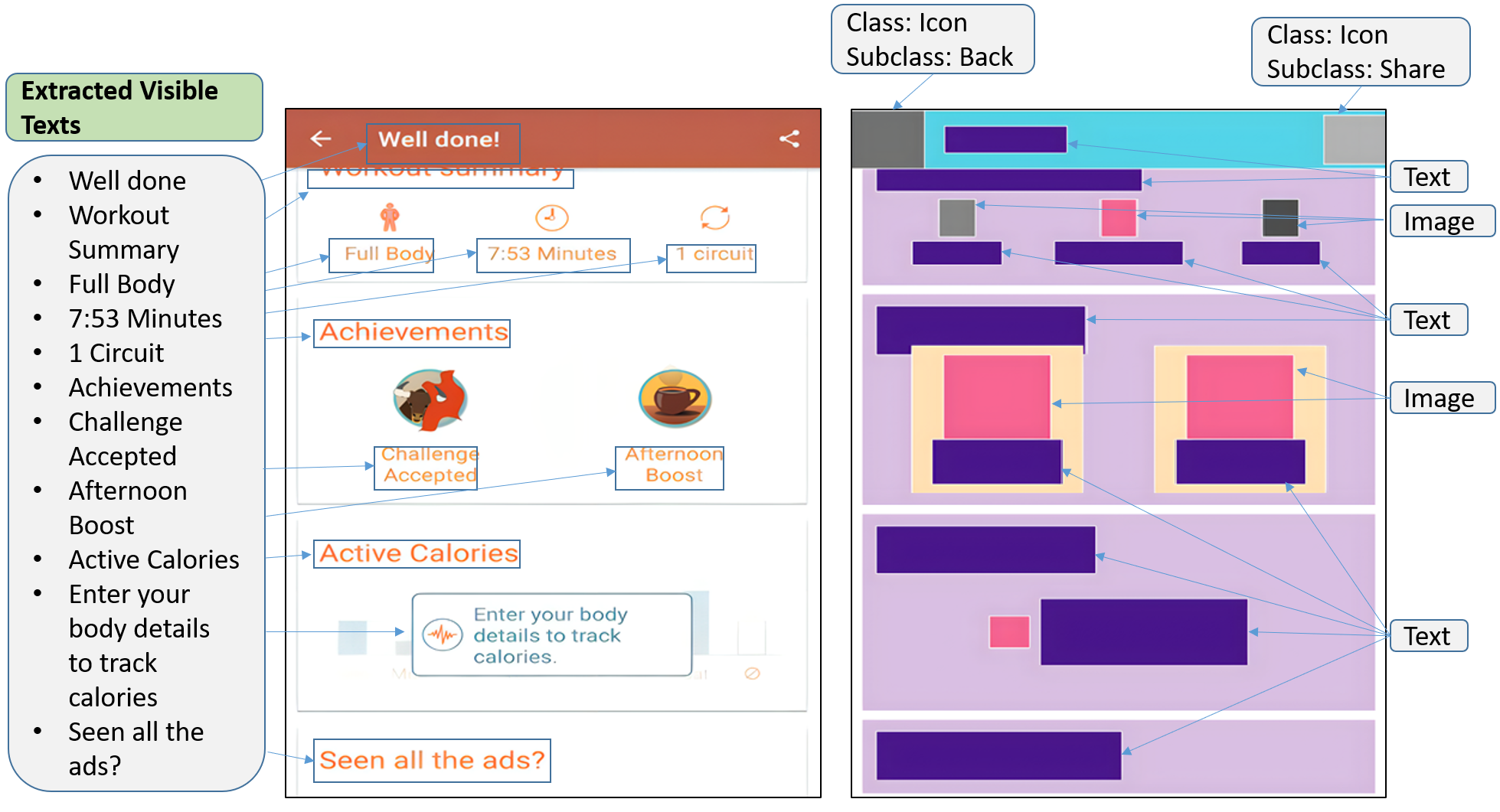}
\caption{The Rico dataset contains all visible text within mobile screens' user interface (UI) hierarchy (depicted on the left). Liu clustered UI elements of Rico mobile screens into 27 categories and sub-categories based on their properties and assigned a descriptive term to represent them (illustrated on the right).
}
\label{fig:semantic_annotaion}
\end{figure*}

Rico, a dataset compiled by Deka et al.~\cite{deka2017rico}, was obtained by mining 72k Android app screens using automatic exploration techniques and crowd-sourcing. Using runtime information of each screen, the researchers extracted the hierarchy of UI elements and attributes of each component, such as its position, Android class, displayed text, etc. Figure~\ref{fig:semantic_annotaion} shows an example Rico mobile screen.

Liu et al.~\cite{liu2018learning} clustered the UI elements of Rico screens based on factors such as image similarity, surrounding text snippets, and code-based patterns. They divided UI elements into 25 categories: ``Text'', ``Icon'', ``Image'', etc. Additionally, they sub-categorized icons into 135~classes and text buttons into 197~types. The study further assigned a description representing the UI element category and its sub-category. Three mobile application developers reach a consensus to select the most appropriate word description representing a specific UI element category to ensure the chosen description's accuracy. Figure~\ref{fig:semantic_annotaion} illustrates examples of word descriptions for the element class and icon sub-category class of a mobile screen.

\subsection{Swire: Search by Full-screen Sketch}

Swire, one of our competing systems, utilizes a complete mobile screen sketch to identify similar Rico mobile screens. Swire collected a 3.8k low-fidelity full-screen Rico screen sketches dataset from four experienced UI designers. They then trained a deep neural network on a subset of this dataset consisting of 1.7k Rico sketch-screenshot pairs. The trained network achieved a top-10 screen retrieval accuracy of 61\%. Swire's approach emulates a traditional paper-based design style, where users sketch with pen on paper inside an Aruco marker frame. 

Each sketch must go through several pipelines, like scanning and adjusting projection. However, changing a sketch using this approach requires starting over, scanning, and following processing steps can be time-consuming. A recent study on Swire has achieved a top-10 accuracy of 90.1\%, currently considered state-of-the-art~\cite{sain2020cross}. Swire sketches follow a predefined drawing convention, where users replace any text with a template "Lorem ipsum dolor" or squiggly lines. Additionally, the deep neural networks employed by Swire are agnostic to the actual content of the text.

\subsection{\toolName{}: Search by Icon Doodles}

Our work is closely related to \toolName{}~\cite{mohian2022psdoodle,mohian2022psdoodle_demo}, an iterative and interactive tool for searching mobile screens from incomplete, partial drawings. \toolName{} supports drawing 24 of the most common Android UI elements. \toolName{} divides the mobile app screen into 24 equal-sized tiles (6 along the width and 4 along the height) to match UI elements in different scales. \toolName{} uses the UI element hierarchy of 58k Rico mobile screens to construct a dictionary that maps each element type to a list of mobile screens. Each screen in the dictionary lists the percentage of the tile area covered by how many instances of that UI element type. With the help of this dictionary, \toolName{} can search through 58k Rico screens and find the best match in real-time.

\toolName{}'s deep neural network trained on \dataName{}~\cite{soumikmohian_DoodleUINet} and \QD{}~\cite{google_quickdraw} recognizes UI element category from drawing strokes.  When the user indicates correct recognition of a UI element, \toolName{}'s algorithm retrieves corresponding Rico screens information from its dictionary. Then assigns a score based on the UI element's type, position, and shape. \toolName{} sorts the mobile app screens by the score to find the top match. 

The \toolName{} website is hosted on AWS and retrieves the top-match mobile screens with a delay of 2 seconds. With a user evaluation,  \toolName{} reported top-10 accuracy of 88.2\% and an average query session of 105s. \toolName{} visual language allows placing a common placeholder for text (squiggle line), not considering any text that appears on the screen. The top-10 accuracy of \toolName{}  is on par with the current state-of-the-art. Nonetheless, in two out of 34 evaluation sessions, the \toolName{} failed to display the target screen on the website.

\section{Overview and Design}

\begin{figure}[h!t]
 \centering
 \includegraphics[width=.9\linewidth,trim={.3in .3in .3in .3in},clip] {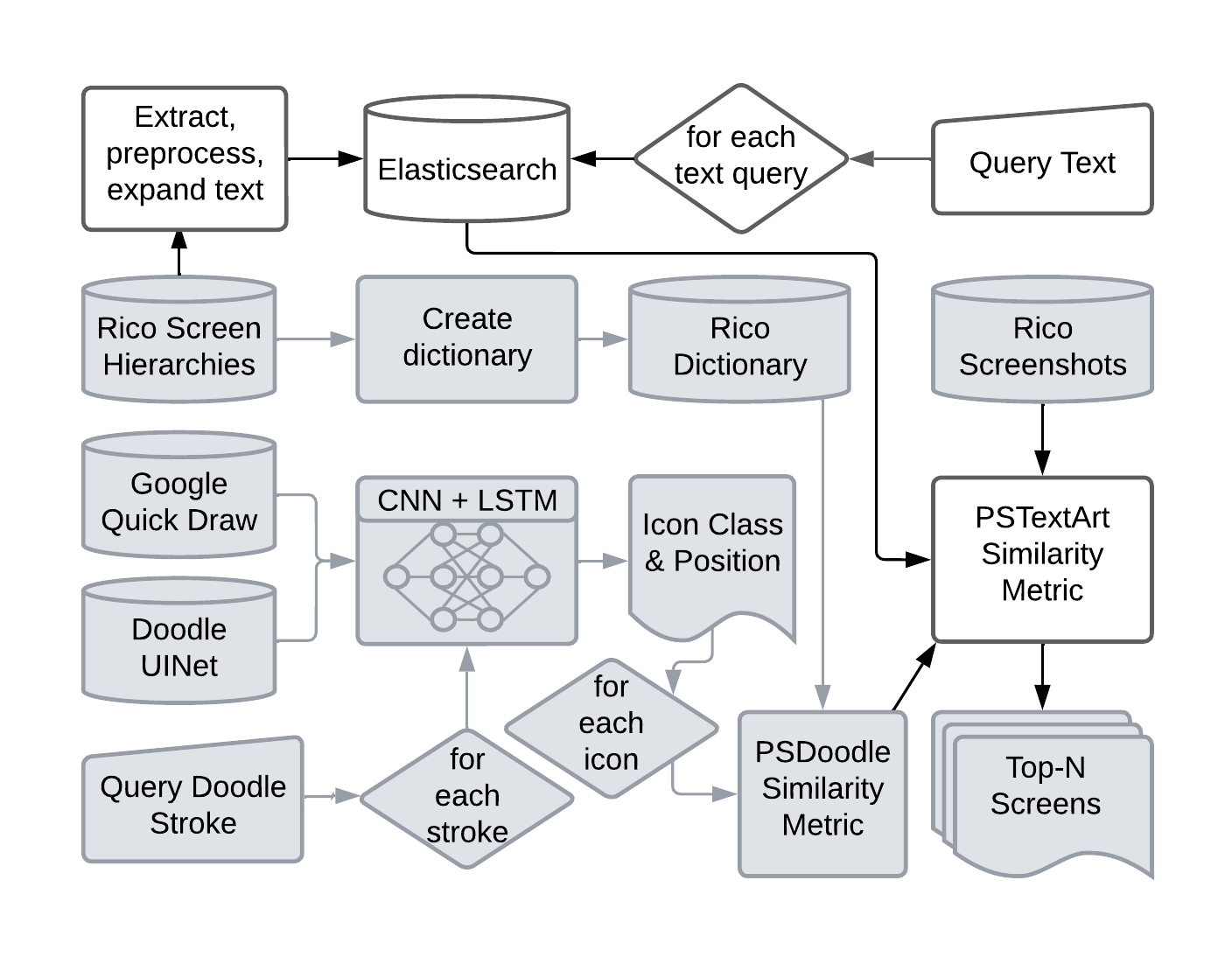}
 \caption{\toolwithtext{} architecture (grayed parts adapted form \toolName{}): To search, a user submits a series of text and icon doodle queries on the \toolwithtext{} site (\url{http://3.137.70.170/WnD/}), which communicates with its back-end on AWS.}
 \label{fig:PSDoodleTextArchitecture}
\end{figure}

\begin{figure}[h!t]
 \centering
 \includegraphics[width=\linewidth,trim={.3in .3in .3in .7in},clip,]
 {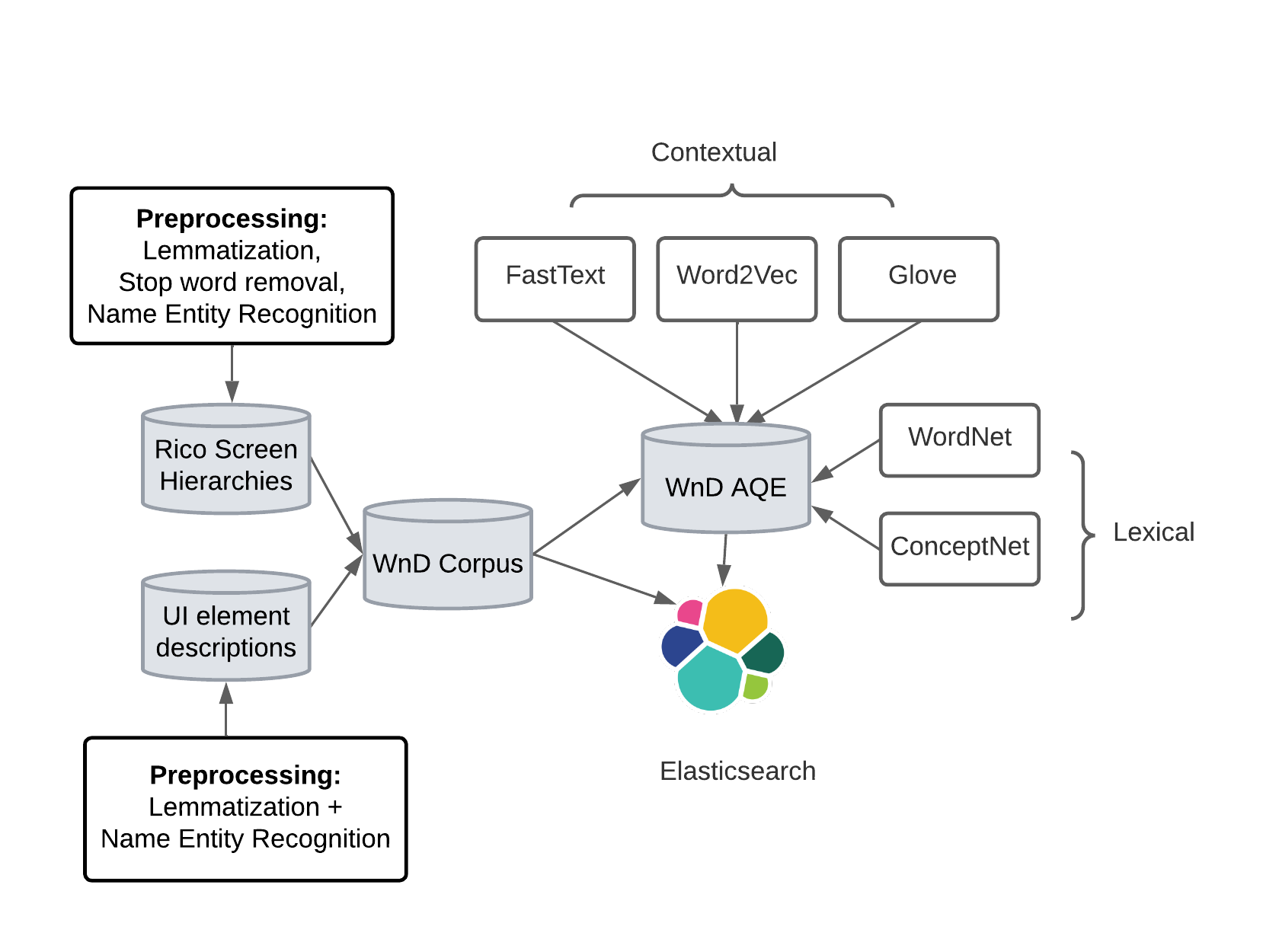}

  \caption{\toolwithtext{} extracts visible texts from Rico's screen hierarchy and UI element description from Liu's work. Each text then goes through preprocessing pipelines to create the \toolwithtext{} corpus. \toolwithtext{} uses lexical and contextual analysis to find each word's top 3 extended terms and store them in the Elasticsearch database. Elasticsearch stores all text indexed by Rico's mobile screen id and position for faster retrieval.}
  \label{fig:Text_Process_Store}
\end{figure}

Figure~\ref{fig:PSDoodleTextArchitecture} provides an overview of \toolwithtext{}'s architecture. \toolwithtext{} adapts \toolName{}'s architecture (Figure~\ref{fig:PSDoodleTextArchitecture}'s grayed elements) to rank a corpus of mobile app screens by their similarity to a query of sketched icons. \toolwithtext{} adds to that by collecting from the Rico dataset~\cite{deka2017rico} mobile screens' visible text and via Liu's work~\cite{liu2018learning} textual descriptions of screens' UI elements. The extracted texts undergo multiple preprocessing steps to improve retrieval performance (Figure~\ref{fig:Text_Process_Store}). Finally, \toolwithtext{} uses Elasticsearch~\cite{elasticsearch} to index the texts for faster retrieval. \toolwithtext{} calculates screen scores for text and sketch queries separately, followed by normalization and merging of these scores to determine the top-N matching screens. 

\subsection{Screen Search Via Icon Doodles \& Positional Text Queries}

\begin{figure}[h!t]
 \centering
 \includegraphics[width=\linewidth,trim={0 0 0 0},clip, width=0.8\linewidth]
 {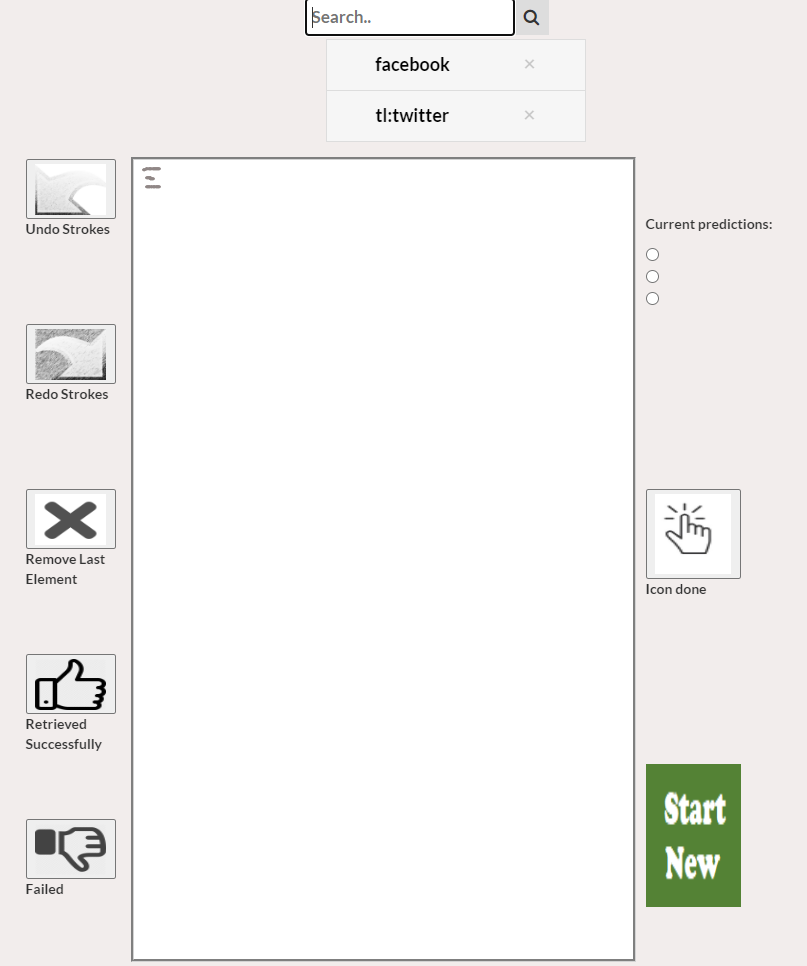}
  \caption{Query interface of \toolwithtext{} web site: Text input area (top) with two submitted queries (``facebook'' and ``tl:twitter'') and sketch input area (bottom) with one submitted query (menu icon). The website also displays \toolwithtext{}'s current top-N Android search result screens (cut from screenshot).}
  \label{fig:searchInterface_Text}
\end{figure}

Figure~\ref{fig:searchInterface_Text} shows the website of \toolwithtext{}. \toolwithtext{} displays the current top-3 predictions for UI elements (top right) when a user draws on the canvas. Users can add or remove text in the query set through the search bar above the canvas in \toolwithtext{}. In the Figure~\ref{fig:searchInterface_Text} example, the user has already issued two text queries, ``facebook'' and ``tl:twitter''.

\begin{figure}[h!t]
 \centering
 \includegraphics[width=\linewidth,clip] {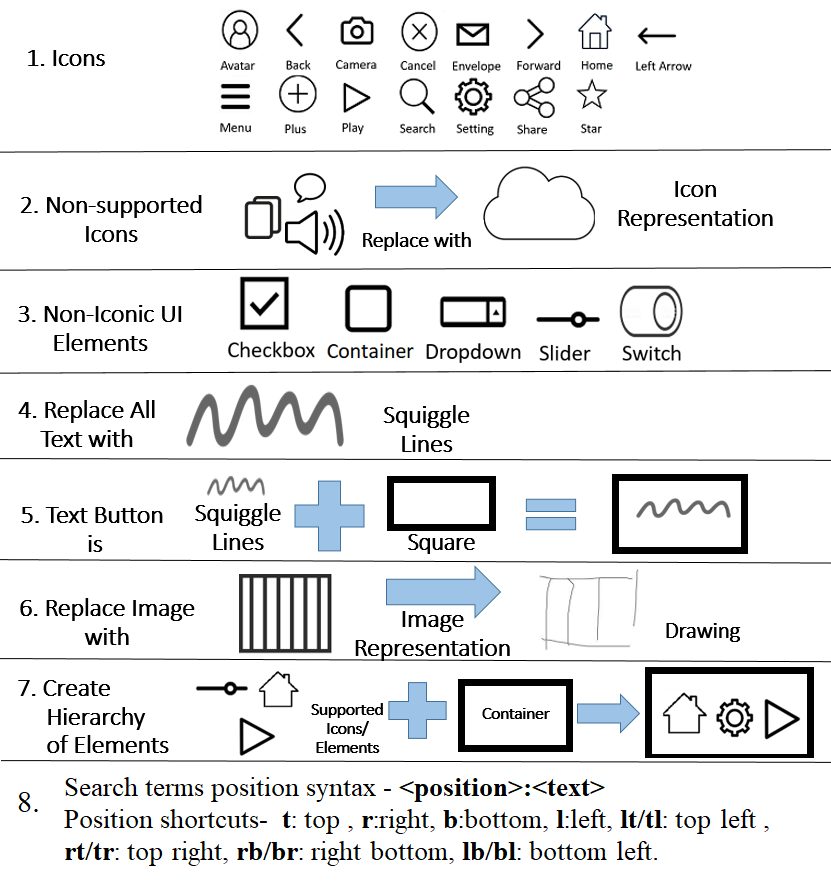}
 \caption{\toolwithtext{} describes its query language to users via this on-screen ``cheat sheet''. 
}
 \label{fig:CheatSheet_Text}
\end{figure}

Figure~\ref{fig:CheatSheet_Text} is the query language of \toolwithtext{} represented to the user as a cheatsheet on its website. Users can draw any atomic UI element, or combine them to create a complex UI element and establish a hierarchy, per the instructions provided in Figure~\ref{fig:CheatSheet_Text}. Upon finishing the current UI element drawing on the canvas, a user signals the tool by clicking the ``Icon done'' button. 

To allow finer-grained text queries, \toolwithtext{} allows the user to specify where a given query text should appear on a screen. This enables users to label text with its position on the screen, facilitating a more precise and targeted search and yielding more accurate results. \toolwithtext{} currently allows the user to specify one of four screen quadrants, top-left (via prefix ``tl:'' or ``lt:''), top-right (``tr:''/``rt:''), bottom-left (``bl:''/``lb:''), and bottom-right (``br:''/``rb:''). Alternatively, \toolwithtext{} allows the user to specify the combination of two adjacent quadrants, i.e., top (``t:''), bottom (``b:''), right (``r:''), or left (``l:''), yielding a total of 12~positional keywords for text queries.

\subsection{Screen Texts, UI Descriptions, And X/Y Pixel Coordinates}

We extract two text types from each of the 58k Rico mobile app screens. First, we parse the screen's hierarchical UI container tree structure to extract all text displayed to the user. On the left of the Figure~\ref{fig:semantic_annotaion} example, this would include the strings listed in the left side-bar, such as ``Well done!'' and ``Full Body''.

Second, from the same hierarchical UI container tree we extract all UI elements and their descriptions according to the similarity-based clustering by Liu et al.~\cite{liu2018learning}. On the right of Figure~\ref{fig:semantic_annotaion}, examples of this type of text describing the screen are in the call-out boxes such as ``Back'' and ``Share''. 

Together with both types of text, we also extract from the hierarchical UI container tree the x/y pixel coordinate location of where the text or UI element appears on the screen. While earlier work has already inferred these texts and on-screen locations for the 58k Rico screens, applying the same techniques to further third-party apps would be straightforward.

\subsection{Screen Contents' Stop Words, Names, And Lemmatization}

To support a wide range of possible user text queries, \toolwithtext{} uses two separate text preprocessing pipelines, one for screen text and the other for UI element descriptions. The involved preprocessing steps are used to improve performance in many natural language processing (NLP) use cases (e.g., in sentiment analysis, document categorization, and document retrieval for user queries~\cite{manning1999foundations, sebastiani2002machine}).

The main difference between \toolwithtext{}'s two preprocessing pipelines is the treatment of stop-words (which in general carry little distinctive information~\cite{abadi2016tensorflow}). While \toolwithtext{} identifies and removes stop-words from on-screen text using the Natural Language Toolkit~\cite{nltk}, \toolwithtext{} preserves stop-words in the UI element descriptions. There stop-words such as ``up'' and ``down''  have distinct meaning that differentiates one UI element from another (e.g., arrow up vs. arrow down).

A simple way to support a wide variety of user queries is to rewrite both the corpus and query via stemming~\cite{lovins1968development,porter1980algorithm}, which using rules maps each word to its root form (e.g., ``walked'' or ``walking'' to ``walk''). To also consider the word's context and part of speech within a sentence, \toolwithtext{} instead maps a word to its morphological root via lemmatization \cite{balakrishnan2014stemming}. This yields a more precise root form, such as ``bet'', or the retention of the original form, such as ``better'', depending on the word's part of speech and context. 

\toolwithtext{} uses spaCy's en\_core\_web\_trf-3.4.1 English language model~\cite{spaCy}, which has excellent accuracy in identifying a word's part of speech in a sentence (precision 0.98), segmenting sentences (precision 0.96), and recognizing named entities (precision 0.9)~\cite{spaCy_accuracy}. \toolwithtext{} stores a word's lemma form in lowercase and consults spaCy to determine if a word is a named entity (such as a person, organization, or location). \toolwithtext{}'s repository includes the resulting preprocessed texts for the 58k Rico screens.

\subsection{Adding Screen Content Synonyms}

While lemmatization covers a variety of user queries, it does not cover synonyms. We thus add the synonyms of our screen contents by combining automatic query expansion~(AQE) approaches. For background, AQE techniques include lexical and contextual analysis, relevance feedback, and supervised approaches~\cite{carpineto2012_AQE_survey}. As relevance feedback and supervised methods rely on user feedback and search history, and we are unaware of public corpora on mobile screen content queries, \toolwithtext{} focuses on lexical and contextual analysis. Lexical approaches encode basic thesaurus-like word relations (e.g., synonym, hypernym, and hyponym). 

Contextual analysis infers such word relations from features such as two words tending to appear together in a large variety of documents. The contextual analysis employs unsupervised deep learning on a large corpus of unstructured text and typically encodes each word's relations to other words as a fixed-length vector of real numbers (aka word embedding)~\cite{word2vec,fasttext}, positioning contextually related words close to each other in vector space~\cite{word2vec, glove}. Then word similarity can be measured via vector distance metrics (e.g.,  l1, l2, and cosine).

To adapt existing contextual models to mobile app screen contents, we tried fine-tuning the pre-trained Word2vec~\cite{word2vec} and GloVe\cite{glove} word embedding models with the local context of our preprocessed Rico mobile app screen contents. However, we quickly realized this not to work well, as each Rico screen on average only contains four words and the screen contents are often disjoint. (e.g., while in close proximity of each other on the Figure~\ref{fig:semantic_annotaion} screen, the two text snippets ``enter your body details to track calories'' and ``seen all the ads?'' represent two different app features). For instance, this yielded a fine-tuned Word2vec~\cite{word2vec} model listing the two most common related words of ``avatar'' as ``edit'' and ``too''.

Word2vec, GloVe, and FastText~\cite{fasttext} represent a variety of contextual approaches and training sets. Word2vec (word2vec-google-news-300 of 3M words and phrases trained on 100B words from Google News) focuses on language-specific information, GloVe (glove-wiki-gigaword-300 of 400k words trained on 5.6B tokens from Wikipedia 2014 + Gigaword) uses global statistical data, and FastText (fasttext-wiki-news-subwords-300 of 1M words trained on 9B+ words) is reliable for out-of-corpus words~\cite{embedding_comparison}. To minimize bias and improve word coverage, \toolwithtext{} combines all three models as follows.

First, for each screen content word, we retrieve the 10~most-similar words from each of the three models via cosine similarity (as cosine considers vector orientation and provides better matches than other metrics such as Euclidean distance~\cite{mikolov2013efficient}). We thus retrieve three sets of up to 10~synonyms each, each synonym with its similarity [0..1]. We then merge the three sets, adding overlapping words' similarity scores and thus favoring common synonyms. 

In the resulting ranked list, we break ties via our two lexical models WordNet~\cite{Wordnet} and ConceptNet~\cite{conceptnet}, demoting a word that is not in WordNet, and to break a further tie demoting a word that is not in ConceptNet. Following earlier work~\cite{hsu2008combining}, we use WordNet~\cite{Wordnet} as it covers English well and works well for query expansion~\cite{fang2008re,hsu2006query} and ConceptNet for its knowledge graph of 5.6M nodes and 34M+ edges.

If a word is not in any of the three contextual models, we again fall back on our lexical models and first look up the word's synonyms in WordNet and then in ConceptNet. This procedure has yielded three synonyms for each word in or describing the 58k~Rico screens. The  exceptions are the 7,448 unnamed entities, as users are typically interested in the entity itself and not a similar entity. The resulting ``Synonym.txt'' file of 29,503~screen content words and their 88,509~synonyms is in \toolwithtext{}'s repository.

\subsection{Storing Screen Contents' Locations And Synonyms}

To support positional text queries, \toolwithtext{} divides each screen into a 2x2 grid of equal-sized rectangles and then maps each extracted screen text or UI element description to one of these four quadrants. While the current \toolwithtext{} implementation maps a screen content's top-left corner's x/y coordinate, this could easily be generalized to mapping the screen content to each quadrant it overlaps with.

\toolwithtext{} stores and indexes its preprocessed screen contents and their on-screen locations together in Elasticsearch. Elasticsearch is widely used for data storage and search~\cite{elasticsearch,divya_elasticsearch,zamfir_elasticsearch}. Specifically, for each word on each Rico screen, \toolwithtext{} maps a pair (words and the phrase's screen quadrant (aka document zone)) to the screen's unique~ID. For each of the (pre-processed) words, \toolwithtext{} further updates Elasticsearch's synonym mapping from that word's pre-processed form to the three synonyms we inferred earlier.

We further configure Elasticsearch to match each query word with Levenshtein edit distance one (i.e., allowing one one-character change per query word). For instance, screen content word ``setting'' would match when inserting (e.g., ``settiing''), deleting (e.g., ``seting''), or replacing (e.g., ``setling'') one character. Choosing edit distance one is a trade-off between allowing more query typos vs. minimizing query response time. \toolwithtext{} runs its website and Elasticsearch on the same AWS EC2 general-purpose computing instance (t2.large: two virtual CPUs and 8 GB of RAM).

\subsection{Drawing Recognition With \toolwithtext{}}

\begin{table*}[htbp]
\begin{center}
  \caption{Doodle recognition trained on 80\% classifying the other 20\% (the test doodles): 
  1st stroke at which a tool ranks a doodle's correct class first (top-1);
  all/fin~=~test doodles' correct class not ranked first in each/final stroke;
  cnt~=~count;
  avg~=~average;
  m~=~median;
  l~=~min;
  h~=~max;
  std~=~standard deviation.
  }
  
  \label{tab:strokeStats_text}
  \begin{tabular}{l|l|rrrrr|rr|rrrrr|rr}
   \hline
    \multicolumn{1}{l|} {\textbf{Category}} &
    \multicolumn{1}{l|} {\textbf{20\%}} &
    \multicolumn{7}{c|} {\textbf{\toolName{} (non-uniform sampling)}} &
    \multicolumn{7}{c} {\textbf{\toolwithtext{} (after re-sampling)}}  
    \\
    \multicolumn{1}{l|} {} &
    \multicolumn{1}{l|} {cnt} &
    \multicolumn{5}{c} {1st stroke to top-1} &
    \multicolumn{2}{c|} {cnt} &
    \multicolumn{5}{c} {1st stroke to top-1} &
    \multicolumn{2}{c} {cnt}
    \\
    & & avg & m & l & h & std & all & fin & avg & m & l & h & std & all & fin \\
      \hline

Camera  &  143  &  2.40 & 2 & 1 & 5 & .77 & 7 & 10  &  2.61 & 2 & 1 & 5 & .85 & 1 & 6  \\ 
Cloud  &  154  &  1.06 & 1 & 1 & 2 & .23 & 2 & 9  &  1.07 & 1 & 1 & 2 & .26 & 3 & 5  \\ 
Envelope  &  145  &  1.87 & 2 & 1 & 6 & .87 & 0 & 7  &  1.89 & 2 & 1 & 6 & .89 & 1 & 4  \\ 
House  &  135  &  2.13 & 2 & 1 & 5 & .98 & 0 & 1  &  2.25 & 2 & 1 & 9 & 1.22 & 3 & 7  \\ 
Jail-window  &  143  &  3.34 & 3 & 1 & 8 & 1.11 & 0 & 7  &  2.58 & 2 & 1 & 7 & 1.02 & 0 & 5  \\ 
Square  &  147  &  1.12 & 1 & 1 & 3 & .34 & 0 & 2  &  1.12 & 1 & 1 & 4 & .42 & 1 & 9  \\ 
Star  &  148  &  1.15 & 1 & 1 & 6 & .57 & 0 & 2  &  1.12 & 1 & 1 & 3 & .39 & 1 & 3  \\ 
    \hline
Avatar  &  136  &  2.87 & 3 & 1 & 7 & .92 & 2 & 3  &  2.79 & 3 & 1 & 5 & .71 & 1 & 3  \\ 
Back  &  122  &  1.06 & 1 & 1 & 3 & .27 & 2 & 3  &  1.02 & 1 & 1 & 3 & .18 & 2 & 1  \\ 
Cancel  &  127  &  1.96 & 2 & 1 & 3 & .57 & 1 & 1  &  1.93 & 2 & 1 & 3 & .44 & 0 & 1  \\ 
Checkbox  &  134  &  2.24 & 2 & 1 & 6 & 1.04 & 2 & 1  &  2.30 & 2 & 1 & 6 & 1.15 & 0 & 2  \\ 
Drop-down  &  133  &  2.49 & 2 & 1 & 9 & 1.19 & 7 & 5  &  3.40 & 3 & 1 & 8 & 1.21 & 3 & 7  \\ 
Forward  &  122  &  1.05 & 1 & 1 & 2 & .22 & 0 & 3  &  1.04 & 1 & 1 & 2 & .20 & 0 & 3  \\ 
Left arrow  &  123  &  1.96 & 2 & 1 & 3 & .56 & 0 & 7  &  1.89 & 2 & 1 & 3 & .48 & 0 & 1  \\ 
Menu  &  126  &  2.16 & 2 & 2 & 8 & .63 & 1 & 3  &  2.91 & 3 & 2 & 6 & .44 & 0 & 1  \\ 
Play  &  127  &  2.03 & 2 & 1 & 7 & .93 & 3 & 21  &  2.03 & 2 & 1 & 5 & .84 & 2 & 3  \\ 
Plus  &  120  &  1.43 & 1 & 1 & 3 & .59 & 2 & 7  &  1.39 & 1 & 1 & 4 & .67 & 0 & 6  \\ 
Search  &  122  &  1.89 & 2 & 1 & 5 & .51 & 3 & 2  &  2.02 & 2 & 1 & 5 & .37 & 0 & 1  \\ 
Setting  &  111  &  2.66 & 2 & 1 & 10 & 1.71 & 0 & 3  &  2.93 & 2 & 1 & 27 & 3.11 & 1 & 8  \\ 
Share  &  117  &  2.97 & 3 & 1 & 7 & 1.15 & 3 & 2  &  3.41 & 3 & 2 & 8 & .88 & 1 & 1  \\ 
Slider  &  134  &  1.85 & 2 & 1 & 6 & .70 & 0 & 4  &  1.86 & 2 & 1 & 4 & .62 & 0 & 2  \\ 
Squiggle  &  144  &  1.08 & 1 & 1 & 3 & .29 & 0 & 0  &  1.12 & 1 & 1 & 4 & .53 & 0 & 0  \\ 
Switch  &  137  &  2.59 & 2 & 1 & 7 & 1.20 & 0 & 8  &  2.59 & 2 & 1 & 8 & 1.09 & 1 & 5  \\ 
         \hline
\end{tabular}
\end{center}
\end{table*}

To recognize as a UI element the stroke sequence the user draws with a mouse or touchscreen, \toolwithtext{} uses \toolName{}'s deep neural network architecture~\cite{mohian2022psdoodle}, i.e., a 1-D convolutional neural network (CNN) layer with 48~filters and kernel size~5, followed by a 1-D CNN layer with kernel size~5 and 64~filters, a 1-D CNN layer with kernel size~3 and 96~filters, 3 Bi-LSTM layers, and a fully-connected layer. Each stroke is a sequence of lines and a line is given by its endpoint x/y coordinates.

\toolwithtext{} trained on sketches from 7~icon classes from \QD{} and 16 from~\dataName{}. Unfortunately, \QD{} and \dataName{} collected sketches using different drawing interfaces and thus different point sampling ratios. In other words, \QD{} and \dataName{} would represent the same stroke via different points (and thus lines). These different sampling ratios make it harder for a deep neural net to learn the icon classes' underlying structure, yielding non-optimal icon sketch recognition performance.

Prior work has resampled sketches to make line endpoints equidistant~\cite{namboodiri2004online}. In contrast, for each stroke \toolwithtext{} determines the Euclidean distance between all pairs of line endpoints and uses 20~times the biggest such distance as the stroke's new number of line endpoints. \toolwithtext{} then resamples each stroke to this new line endpoint count using NumPy's one-dimensional linear interpolation. While \toolName{} reported a 94.2\% sketch recognition accuracy, after re-training the network on our resampled sketches for 18,131 steps, \toolwithtext{} achieves 95.8\% accuracy on \toolName{}'s test data. The achieved level of precision mentioned in this study is comparable to that of the \appName{} \cite{Doodle2App}, albeit across a more extensive range of categories.

Table~\ref{tab:strokeStats_text} compares icon doodle recognition performance with and without \toolwithtext{}'s resampling. Although the average 1st stroke at which \toolwithtext{} ranks a doodle's correct class remained largely unchanged, the resampling pipeline reduced the number of misclassifications at both the final and all strokes. For example, while \toolName{} did not rank camera top-1 for 7/143 camera sketches in each stroke and 10/143 in the final stroke, these ratios reduce to 1/143 and 6/143 for \toolwithtext{}. These findings suggest that the addition of a resampling pipeline has a positive impact on \toolwithtext{}'s doodle recognition accuracy.

\subsection{Ranking Mobile Screens}

When the user adds or removes an icon sketch or adds or changes a query text phrase, \toolwithtext{} re-scores each of its 58k~Rico screens by how closely the screen matches the new combination of icon sketches and text queries. First (upper half of Algorithm~\ref{alg:algorithm2}), \toolwithtext{} scores each screen based on how closely it matches the size and location of the doodles of one icon class at a time. \toolwithtext{} retrieves these screen scores from \toolName{} and then normalizes these scores to ensure that each sketch element can contribute equally.

In the second phase, \toolwithtext{} starts processing text queries by first tokenizing each query, where tokens are separated by a non-empty sequence consisting of whitespace (`` ''), dot (``.''), comma (``,''), exclamation mark (``!'''), and question mark (``?'') characters. Within the tokens, \toolwithtext{} then checks for one of 12~positional keywords and maps it to Elasticsearch fields. For example, ``lt:'' or ``tl:'' specifies a screen's top right quadrant. Each of the four more general keywords ``t:'', ``b:'', ``l:'', and  ``r:'' \toolwithtext{} maps to its two underlying quadrants, e.g., ``t:'' to top-left and top-right. Finally, if a text query has no positional keyword, \toolwithtext{} maps the query to all four quadrants. 

\toolwithtext{} then performs basic stemming (removing prefixes or suffixes), converts each word to lowercase, and sends the resulting query to Elasticsearch. Elasticsearch matches the query in the corresponding position fields and returns the resulting screen IDs and their query score. \toolwithtext{} configured Elasticsearch to calculate scores with weights 10 for exact match and 4 for matching a synonym. \toolwithtext{} then scores all screens one text query at a time and adds these (normalizes) scores to the screens' overall scores. Finally, \toolwithtext{} sorts screens by descending scores, yielding the screens' search result ranking.

To offset the normalization effect, we adjusted \toolName{}'s five hyperparameters. To find their optimal values that would result in a high score and top rank for the target screen with the new algorithm, \toolwithtext{} used scikit-learn's GridSearchCV~\cite{sklearn_api} to perform a comprehensive search on the 30~sketches \toolName{} used and published online. The resulting new hyperparameter values are
$p_1=11$,
$p_2=1$,
$p_3=1$, 
$\Delta_w=0.7$, and 
$C_w=12$.

\begin{algorithm}
\caption{\toolwithtext{} calculates the screen score for the doodle and text queries separately, then adds the normalized scores;
sketch~=~list of maps from icon class to doodles of that class;
search(doodles)~=~score screens by how closely they match the size and location of the current class's doodles (via \toolName{}).
}
\label{alg:algorithm2}
\begin{algorithmic}[1] 
\State $res := $ \{\} \Comment{map: screen $\rightarrow$ score}
\For{$doodles$ in $sketch$}  \Comment{iterate over icon classes}
      \State $res_{ddl} := $ \{\} \Comment{map: screen $\rightarrow$ score}
      \State ${res_{ddl}} := search(doodles)$ \Comment{scores for class doodles}
      \For{$screen$ in ${res_{ddl}}$}  \Comment{normalize and update}
        \State $res[screen] +=   \dfrac{{res_{ddl}}[screen]}{max({res_{ddl}})} * len(doodles)$
      \EndFor
    \EndFor

\For{$text$ in $queries$} \Comment{iterate over text queries}
     \State $res_{txt} := $ \{\} \Comment{map: screen $\rightarrow$ score}
      \State ${res_{txt}} := search(text)$ \Comment{scores for text}
      \For{$screen$ in ${res_{txt}}$}   \Comment{normalize and update}
         \State $res[screen] += \dfrac{{res_{txt}}[screen]}{max({res_{txt}})}$
      \EndFor
    \EndFor
\State sort($res$)   \Comment{sort screens by score in descending order} 
\end{algorithmic}
\end{algorithm}

\section{Evaluation}

We evaluated \toolwithtext{} by comparing its top-10 retrieval accuracy, response time, and relevance score with the state-of-the-art approaches \toolName{} and Swire, which are both sketch-only. Specifically, we explore the following research questions to assess \toolwithtext{}'s effectiveness. 

 \begin{description}
 \item[RQ1] At similar tool runtime, can \toolName{} achieve similar top-10 accuracy as state-of-the-art screen search approaches?
 \item[RQ2] At similar top-10 accuracy, can \toolwithtext{} reduce the state-of-the-art approaches' overall runtime?
 \item[RQ3] Does \toolwithtext{} retrieve results relevant to the query consisting of text and UI element sketches?
 \item[RQ4] When given a choice, do \toolwithtext{} users search for a screen via a mix of UI element sketches and text or do they exclusively use one of these styles?
\end{description} 

To measure how well the various approaches perform, we use a top-k retrieval accuracy metric. Top-k retrieval accuracy is a commonly used metric in information retrieval (including by all of our competitor approaches). Top-k retrieval accuracy is also a good indicator of user satisfaction~\cite{huffman2007well}. To measure this accuracy, we randomly selected 26 of the 58k Rico screens that (a) have at least two \toolwithtext{}-supported UI elements and (b) have a corresponding Swire drawing. We then showed a participant such a screen as a target screen and measured how high \toolwithtext{} ranked this target screen in its search results.

To evaluate \toolwithtext{}, we recruited 10 Computer Science students for a user study. These participants had a similar background and qualification as the user study participants conducted to assess the effectiveness of \toolName{}~\cite{mohian2022psdoodle}. Specifically, we selected participants with no formal UI/UX design training. To ensure a diverse group, we recruited five participants with prior mobile app development experience (3~female, 2~male) and five without prior mobile app development experience (2~female, 3~male). Each participant received USD~10 as compensation for their time. For this study, they accessed \toolwithtext{} for the first time via the internet on their own device.

During the study, each participant spent on average 10~minutes going through an interactive tutorial on a website hosted on \url{http://pixeltoapp.com//toolInsVT/}. The tutorial covered the basics of \toolwithtext{}'s visual query language for sketching UI~elements, drawing techniques, and search functionalities of \toolwithtext{}. We then instructed each participant to search for at least three screens, one screen at a time. 

The website recorded the participant's drawings, text queries, the results of their search query, and the total time they spent for each target screen (which includes time for sketching, waiting for tool feedback, and thinking about how to proceed with a search). We observed the participants' performance via screen sharing but did not further guide or influence the participants. 

The instructions directed a participant to keep searching until they found the target screen in the search results or until three minutes passed and we enforced this time limit. Once a participant completed the search for a target screen, we asked the participants to rate how relevant they felt the top 10 search results were to their query. All this information is available in the \toolwithtext{} repository. 
 % \SM{Yes. Same as before}

\begin{table*}[htbp]
\begin{center}
  \caption{A participant(P) iteratively issued s~sketch and x~text queries and spent t[s] time(in seconds) to retrieve a specified target screen using \toolwithtext{}. For most of the session, the target screen's position~(p) was in the top 10. The participant (P) also provided ratings indicating the number of top-10 relevant search results (r) to the query. }
  
  \label{tab:participantsscoretext}
  \begin{tabular}{r|ccrrr|ccrrr|ccrrr}
   \hline
    \multicolumn{1}{l|} {\textbf{P}} &
    \multicolumn{5}{c|} {\textbf{Target 1}} &
    \multicolumn{5}{c|} {\textbf{Target 2}} &
    \multicolumn{5}{c} {\textbf{Target 3}}
    \\
    & s & x & t[s]  & p & r & s & x & t[s] & p & r & s & x & t[s] & p & r   \\
      \hline
1 & 3 & 1 & 96 & 4 & 8 & 2 & 1 & 32 & 4 & 6 & 3 & 2 & 71 & 1 & 4 \\
2 & 2 & 3 & 12 & 1 & 4 & 4 & 1 & 50 & 29 & 0 & 2 & 3 & 156 & 1 & 2 \\
3 & 6 & 0 & 89 & 15 & 6 & 2 & 1 & 23 & 1 & 5 & 3 & 0 & 52 & 2 & 4 \\
4 & 3 & 1 & 125 & 1 & 4 & 3 & 1 & 38 & 2 & 8 & 3 & 0 & 19 & 11 & 10 \\
5 & 2 & 1 & 15 & 3 & 6 & 2 & 1 & 5 & 3 & 2 & 3 & 2 & 60 & 1 & 6 \\
6 & 2 & 1 & 26 & 2 & 3 & 3 & 0 & 38 & 4 & 10 & 2 & 1 & 12 & 7 & 4 \\
7 & 2 & 1 & 41 & 1 & 10 & 2 & 1 & 17 & 4 & 7 & 2 & 2 & 18 & 5 & 5 \\
8 & 2 & 4 & 90 & 1 & 5 & 3 & 2 & 14 & 7 & 6 & 5 & 0 & 57 & 1 & 5 \\
9 & 3 & 2 & 48 & 2 & 3 & 2 & 1 & 28 & 2 & 7 & 2 & 1 & 17 & 4 & 4 \\
10 & 2 & 1 & 30 & 1 & 4 & 3 & 0 & 68 & 3 & 4 & 3 & 1 & 18 & 1 & 4 \\

    \hline
\end{tabular}
\end{center}
\end{table*}

\subsection{RQ1: Achieving state-of-the-art performance in Top-10 Accuracy}

We asked 10 participants to search for 3~target screens each, yielding 30~search sessions. We measured the top-10 retrieval accuracy and found that out of the 30 final queries, \toolwithtext{} displayed the target screen in the top-10 search results 28~times, resulting in a 93.13\% top-10 accuracy. This surpasses the top-10 accuracy of Swire (61\%), its follow-up work (90.1\%), and \toolName{} (88.2\%).

\subsection{RQ2: Efficient and Interactive Screen Retrieval}

We analyzed the time participants searched for a target screen using \toolwithtext{} and compared it with the state-of-the-art tools Swire and \toolName{}. \toolwithtext{} search sessions on average took less time than the other two tools. The average time it took for Swire participants to draw a Rico screen was 246s, while the duration of sketching per search session using \toolName{} ranged from 30s to 259s, with an average of 107s. In comparison, the search time for \toolwithtext{} varied from 5s to 156s with an average of 45s. Table~\ref{tab:participantsscoretext} provides the complete duration of each search session for all 30 experiments, including the final position of the target Rico screen in the search results.  

\toolwithtext{} maintains \toolName{}'s iterative search approach by allowing users to refine their search queries by adding or removing text or sketches. The search process for \toolwithtext{} begins with adding the first text or drawing. Swire requires users to sketch every element on a mobile screen, resulting in the same number of UI elements as a Rico mobile screen. On average, Swire sketches of the 26 target screens had 21.1 UI elements. For query sessions of \toolName{}, the participants sketched an average of 5.5 UI elements. In \toolwithtext{}, the number of text and UI element sketches in a single session was only 3.9. These results suggest that \toolwithtext{} can find the target screen more accurately with less information than its competitors.

Additionally, \toolwithtext{} offers interactive search functionality and is hosted on AWS. We observed during our experiments that \toolwithtext{} updated the top-10 result screens on the user's website in less than 2 seconds from the time the user submitted a search query. Apart from communication with AWS, the primary time factors were sketch recognition (below 0.1 seconds), screen similarity calculation, and screen ranking (below 1 second).

\subsection{RQ3: Displaying Multiple Relevant Screens}

Participants evaluated the quality of \toolwithtext{}'s top-10 result screens after each of 30 search sessions, resulting in the evaluation of a total of 300 screens. Among these, participants judged 156 screens as relevant to their search query. Table \ref{tab:participantsscoretext} displays the study results, indicating that \toolwithtext{} successfully retrieved multiple relevant screens for each query. The relevance score achieved by \toolwithtext{} is 52\%, which surpassed the score of 42\% obtained by \toolName{}. 

As an example of relevant \toolwithtext{} results, each row in the Figure~\ref{fig:iterativeRetriveval_Text} motivating example depicts a query and \toolwithtext{}'s top 5 search results. Most search results accurately position the sketched UI element, and most have text queries on the screen. Overall, the results suggest that \toolwithtext{} is better at retrieving multiple relevant search results.

\subsection{RQ4: Improving Search Relevance via Text \& Sketch}

While \toolwithtext{} surpassed \toolName{} in terms of accuracy and speed, another key advantage of \toolwithtext{} is that it consistently presented the target screen on the first page of 50 search result screens, requiring users to provide fewer query information. In contrast, during \toolName{}'s evaluation, 2 out of 34 participants failed to find the target screen on the first page within 3 minutes of their query session. 

Many of \toolName{}'s shortcomings are due to inaccurate clustering of UI elements in Liu's work. \toolwithtext{} can side-step this inaccurate clustering of UI elements, by enabling users to pinpoint the screen via text queries. As illustrated in the first row of Figure~\ref{fig:fail_scenario}, \toolwithtext{} ranks the target screen within the top 900  but fails to bring the target screen to the first page using only sketch queries due to the misclassification of the ``switch'' icon. However, by adding a single text query, the participant obtained the desired result within the top 30. This observation highlights the utility of incorporating text queries in conjunction with sketch queries to improve the accuracy and efficiency of the system in retrieving relevant results.

 Table~\ref{tab:participantsscoretext} illustrates how users combined text and sketches to retrieve the desired Rico screen. In 30 search sessions, 25 combined text and sketches, while the remaining five solely used drawings.

\subsection{Screen Searches Ranked Outside Top-10}

\begin{figure}[h!t]
 \centering
 \includegraphics[width=\linewidth,trim={0 0 0 0},clip, width=0.8\linewidth]
 {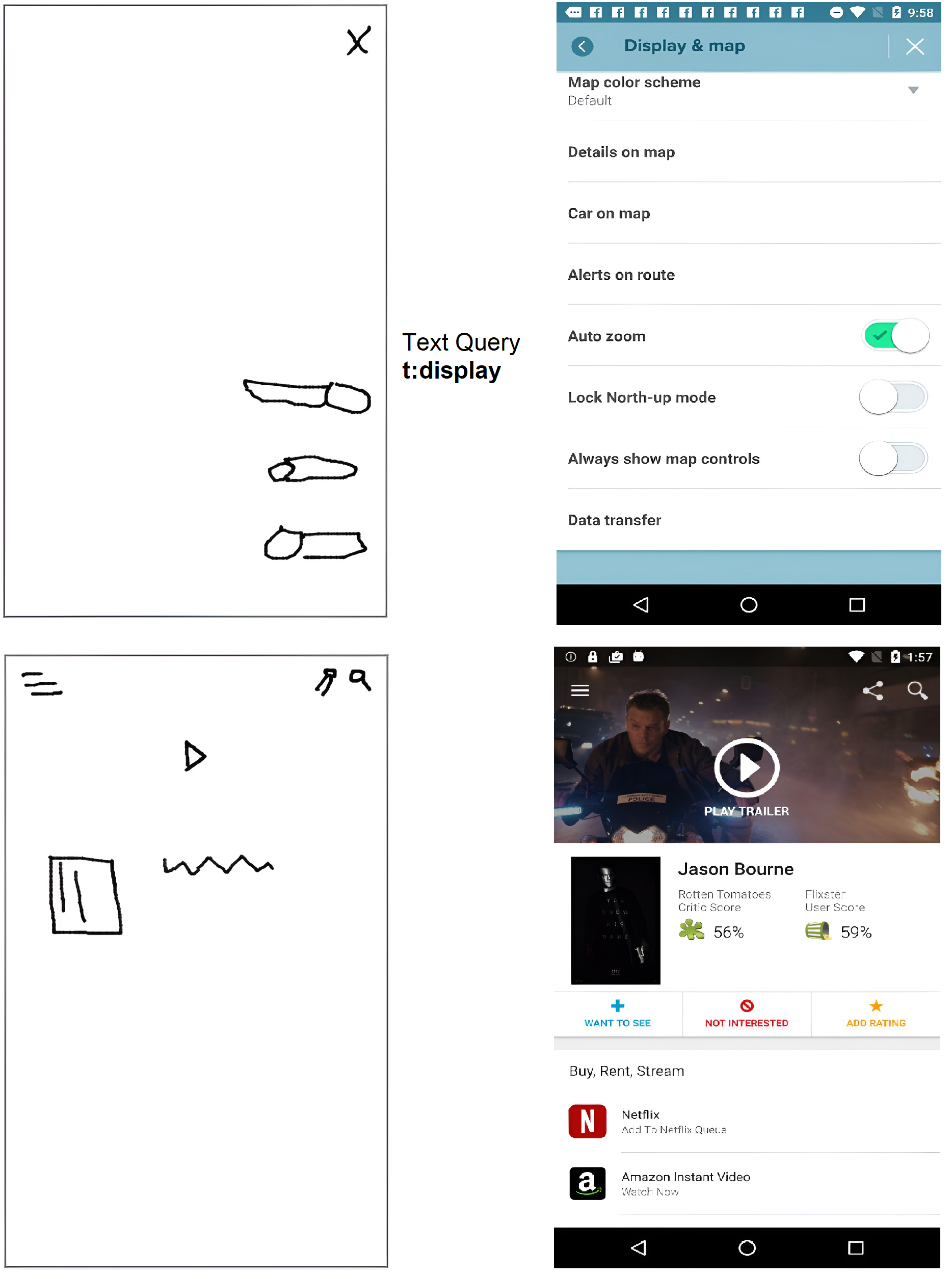}
  \caption{During \toolwithtext{} evaluation study, there were three sessions where the target screen did not appear in the top 10 results. In one case, Participant-2 searched for the target screen (top-right ) by sketching 'close,' three 'switch' icons, and text query 't: display.' In another case, Participant-3 searched the target screen (bottom-right) by sketching  'menu,' 'play,' 'image,' 'share,' 'squiggle,' and 'search.'  The participant stops the search when \toolwithtext{} brings the target screen to the first page of results (top 50).}
  \label{fig:fail_scenario}
\end{figure}

Notably, on three occasions, \toolwithtext{} could not locate the target screen within the top 10 search results. Figure~\ref{fig:fail_scenario} depicts two out of the three search sessions. In one instance, During the second session, Participant-2 searched the target screen (right) by sketching ``close'' and three ``switch'' icons. Initially, the target screen was ranked relatively low, around 900, due to Rico's hierarchy labeling the ``switch'' icon as ``check-in'' for the target screen. However, after the user added a text query "t:display", it was ranked in the top 30. 

In another instance, Participant-3 searched for the target screen by sketching a ``play'' icon, but \toolwithtext{} failed to fetch the target screen because Liu et al. labeled the ``play'' icon as an image. However, when Participant-3 added more search terms including ``menu'', ``image'', ``share'', ``squiggle'', and ``search'', \toolwithtext{} was able to re-rank the target screen as rank 13,900, 16,425, 3,162, 88, and 15. 

In both cases, the participants stopped searching as the target screen appeared on the first page of the search results. The results obtained from the study suggest that \toolwithtext{} can reliably fetch a target screen with high accuracy. All these findings are available in the \toolwithtext{} repository.

\subsection{User Feedback}

Eight of ten participants have responded to our post-experiment survey. The survey contained the following three open-ended questions.

\begin{description}
  \item[Q1] How can we improve your search interface?
  \item[Q2] What was your overall experience with the search results?
  \item[Q3] How was your overall experience?
\end{description}

For the first question, users have provided suggestions to improve interactivity and make minor changes to the user interface. For example, one participant asked us ``[to] enhance the UI for the website''. Another participant left us with the following notable quote.
\begin{quote}
``Refine the Green buttons, iterate through a `not so comic' UI. eg, Green next button.''
\end{quote}

In response to the second question, participants provided feedback regarding the search results and expressed satisfaction with the current output. Following quotes from participants are noteworthy.

\begin{quote}
``They were surprisingly accurate quite quickly, but I don't know how big the data set is.''
\end{quote}

\begin{quote}
``The idea and the accuracy really amazed me.''
\end{quote}

\begin{quote}
``Great, bit of lag with the refresh, but that's ok.''
\end{quote}

Lastly, participants rated their overall experience with the tool on a scale of 1 to 10. Responses ranged from 6 to 10. The users' average and median ratings were 8. These results and the survey's open-ended responses are accessible in \toolwithtext{}'s repository.

\section{Conclusions}

The process of locating a particular mobile app screen in existing repositories is conventionally restricted to keyword searches or necessitates the creation of either a complete sketch, as seen in Swire, or a partial sketch, as seen in PSDoodle. \toolwithtext{} is the first approach that combines text and drawing input to search for mobile screens, offering an interactive and iterative search engine. \toolwithtext{} developed using a combination of the Rico repository, which contains approximately 58k Android app screens, the Google \QD{} dataset comprising icon-level doodles, and \dataName{} curated corpus of some 10,000 app icon doodles. In our assessment involving software developers external to our team, \toolwithtext{} outperformed existing interactive searching tools by providing superior top-10 search accuracy with considerably less time required to complete the search. Thus, \toolwithtext{} constitutes a significant advancement in the mobile app screen search field, and it is available for use by interested parties under open-source licenses.

\bibliographystyle{IEEEtran}
\bibliography{ref}

% Generated by IEEEtran.bst, version: 1.12 (2007/01/11)
\begin{thebibliography}{10}
\providecommand{\url}[1]{#1}
\csname url@samestyle\endcsname
\providecommand{\newblock}{\relax}
\providecommand{\bibinfo}[2]{#2}
\providecommand{\BIBentrySTDinterwordspacing}{\spaceskip=0pt\relax}
\providecommand{\BIBentryALTinterwordstretchfactor}{4}
\providecommand{\BIBentryALTinterwordspacing}{\spaceskip=\fontdimen2\font plus
\BIBentryALTinterwordstretchfactor\fontdimen3\font minus
  \fontdimen4\font\relax}
\providecommand{\BIBforeignlanguage}[2]{{%
\expandafter\ifx\csname l@#1\endcsname\relax
\typeout{** WARNING: IEEEtran.bst: No hyphenation pattern has been}%
\typeout{** loaded for the language `#1'. Using the pattern for}%
\typeout{** the default language instead.}%
\else
\language=\csname l@#1\endcsname
\fi
#2}}
\providecommand{\BIBdecl}{\relax}
\BIBdecl

\bibitem{bernal2019guigle}
C.~Bernal-C{\'a}rdenas, K.~Moran, M.~Tufano, Z.~Liu, L.~Nan, Z.~Shi, and
  D.~Poshyvanyk, ``Guigle: A gui search engine for android apps,'' in
  \emph{2019 IEEE/ACM 41st International Conference on Software Engineering:
  Companion Proceedings (ICSE-Companion)}.\hskip 1em plus 0.5em minus
  0.4em\relax IEEE, 2019, pp. 71--74.

\bibitem{kolthoff2023data}
K.~Kolthoff, C.~Bartelt, and S.~P. Ponzetto, ``Data-driven prototyping via
  natural-language-based gui retrieval,'' \emph{Automated Software
  Engineering}, vol.~30, no.~1, p.~13, 2023.

\bibitem{mohian2022psdoodle}
S.~Mohian and C.~Csallner, ``Psdoodle: fast app screen search via partial
  screen doodle,'' in \emph{Proceedings of the 9th IEEE/ACM International
  Conference on Mobile Software Engineering and Systems}, 2022, pp. 89--99.

\bibitem{huang2019swire}
F.~Huang, J.~F. Canny, and J.~Nichols, ``Swire: Sketch-based user interface
  retrieval,'' in \emph{Proceedings of the 2019 CHI Conference on Human Factors
  in Computing Systems}.\hskip 1em plus 0.5em minus 0.4em\relax ACM, May 2019.

\bibitem{sain2020cross}
A.~Sain, A.~K. Bhunia, Y.~Yang, T.~Xiang, and Y.-Z. Song, ``Cross-modal
  hierarchical modelling for fine-grained sketch based image retrieval,'' in
  \emph{Proc. 31st British Machine Vision Virtual Conference (BMVC)}, 2020.

\bibitem{ines2017evalmobileinterface}
G.~Ines, S.~Makram, C.~Mabrouka, and A.~Mourad, ``Evaluation of mobile
  interfaces as an optimization problem,'' \emph{Procedia computer science},
  vol. 112, pp. 235--248, 2017.

\bibitem{hellmann2011ruletestUI}
T.~D. Hellmann and F.~Maurer, ``Rule-based exploratory testing of graphical
  user interfaces,'' in \emph{2011 Agile Conference}.\hskip 1em plus 0.5em
  minus 0.4em\relax IEEE, 2011, pp. 107--116.

\bibitem{user_interface_growth_2028}
\BIBentryALTinterwordspacing
``User interface (ui) design market size, share, growth, and industry analysis,
  by type(user experience (ux) design, interaction design (id), visual \&amp;
  graphic design and others), by application(software and app, web page, game,
  tv interfaces and others), regional forecast to 2028).'' [Online]. Available:
  \url{https://www.businessresearchinsights.com/market-reports/user-interface-ui-design-market-102500}
\BIBentrySTDinterwordspacing

\bibitem{deka2017rico}
B.~Deka, Z.~Huang, C.~Franzen, J.~Hibschman, D.~Afergan, Y.~Li, J.~Nichols, and
  R.~Kumar, ``Rico: {A} mobile app dataset for building data-driven design
  applications,'' in \emph{Proc. 30th Annual {ACM} Symposium on User Interface
  Software and Technology (UIST)}.\hskip 1em plus 0.5em minus 0.4em\relax ACM,
  Oct. 2017, pp. 845--854.

\bibitem{mohian2022psdoodle_demo}
S.~Mohian and C.~Csallner, ``Psdoodle: searching for app screens via
  interactive sketching,'' in \emph{Proceedings of the 9th IEEE/ACM
  International Conference on Mobile Software Engineering and Systems}, 2022,
  pp. 84--88.

\bibitem{liu2018learning}
T.~F. Liu, M.~Craft, J.~Situ, E.~Yumer, R.~Mech, and R.~Kumar, ``Learning
  design semantics for mobile apps,'' in \emph{Proc. 31st Annual ACM Symposium
  on User Interface Software and Technology (UIST)}, 2018, pp. 569--579.

\bibitem{soumikmohian_DoodleUINet}
\BIBentryALTinterwordspacing
S.~Mohian and C.~Csallner, ``{DoodleUINet: Repository for DoodleUINet Drawings
  Dataset and Scripts},'' Jul. 2021. [Online]. Available:
  \url{https://doi.org/10.5281/zenodo.5144472}
\BIBentrySTDinterwordspacing

\bibitem{google_quickdraw}
\BIBentryALTinterwordspacing
``Google quickdraw.'' [Online]. Available:
  \url{https://quickdraw.withgoogle.com/}
\BIBentrySTDinterwordspacing

\bibitem{elasticsearch}
B.~Elasticsearch, ``Elasticsearch,'' \emph{Internet: https://www.
  elastic.co,[Nov. 21, 2022]}, 2018.

\bibitem{manning1999foundations}
C.~Manning and H.~Schutze, \emph{Foundations of statistical natural language
  processing}.\hskip 1em plus 0.5em minus 0.4em\relax MIT press, 1999.

\bibitem{sebastiani2002machine}
F.~Sebastiani, ``Machine learning in automated text categorization,'' \emph{ACM
  computing surveys (CSUR)}, vol.~34, no.~1, pp. 1--47, 2002.

\bibitem{abadi2016tensorflow}
M.~Abadi \emph{et~al.}, ``Tensorflow: {A} system for large-scale machine
  learning,'' in \emph{Proc. OSDI}.\hskip 1em plus 0.5em minus 0.4em\relax
  USENIX, Nov. 2016, pp. 265--283.

\bibitem{nltk}
S.~Bird, E.~Klein, and E.~Loper, \emph{Natural language processing with Python:
  analyzing text with the natural language toolkit}.\hskip 1em plus 0.5em minus
  0.4em\relax " O'Reilly Media, Inc.", 2009.

\bibitem{lovins1968development}
J.~B. Lovins, ``Development of a stemming algorithm.'' \emph{Mech. Transl.
  Comput. Linguistics}, vol.~11, no. 1-2, pp. 22--31, 1968.

\bibitem{porter1980algorithm}
M.~F. Porter, ``An algorithm for suffix stripping,'' \emph{Program}, 1980.

\bibitem{balakrishnan2014stemming}
V.~Balakrishnan and E.~Lloyd-Yemoh, ``Stemming and lemmatization: A comparison
  of retrieval performances,'' 2014.

\bibitem{spaCy}
M.~Honnibal, I.~Montani, S.~Van~Landeghem, and A.~Boyd, ``{spaCy}:
  Industrial-strength natural language processing {(NLP)} in {Python},'' 2020.

\bibitem{spaCy_accuracy}
\BIBentryALTinterwordspacing
explosion\_ai, ``Release en\_core\_web\_trf-3.4.1 · explosion/spacy-models.''
  [Online]. Available:
  \url{https://github.com/explosion/spacy-models/releases/tag/en_core_web_trf-3.4.1}
\BIBentrySTDinterwordspacing

\bibitem{carpineto2012_AQE_survey}
C.~Carpineto and G.~Romano, ``A survey of automatic query expansion in
  information retrieval,'' \emph{Acm Computing Surveys (CSUR)}, vol.~44, no.~1,
  pp. 1--50, 2012.

\bibitem{word2vec}
T.~Mikolov, K.~Chen, G.~Corrado, and J.~Dean, ``Efficient estimation of word
  representations in vector space,'' \emph{arXiv preprint arXiv:1301.3781},
  2013.

\bibitem{fasttext}
P.~Bojanowski, E.~Grave, A.~Joulin, and T.~Mikolov, ``Enriching word vectors
  with subword information,'' \emph{Transactions of the association for
  computational linguistics}, vol.~5, pp. 135--146, 2017.

\bibitem{glove}
J.~Pennington, R.~Socher, and C.~D. Manning, ``Glove: Global vectors for word
  representation,'' in \emph{Proceedings of the 2014 conference on empirical
  methods in natural language processing (EMNLP)}, 2014, pp. 1532--1543.

\bibitem{embedding_comparison}
E.~M. Dharma, F.~L. Gaol, H.~Leslie, H.~Warnars, and B.~Soewito, ``The accuracy
  comparison among word2vec, glove, and fasttext towards convolution neural
  network (cnn) text classification,'' \emph{J Theor Appl Inf Technol}, vol.
  100, no.~2, p.~31, 2022.

\bibitem{mikolov2013efficient}
T.~Mikolov, K.~Chen, G.~Corrado, and J.~Dean, ``Efficient estimation of word
  representations in vector space,'' \emph{arXiv preprint arXiv:1301.3781},
  2013.

\bibitem{Wordnet}
C.~Fellbaum, ``Wordnet and wordnets,'' 2005.

\bibitem{conceptnet}
R.~Speer, J.~Chin, and C.~Havasi, ``Conceptnet 5.5: An open multilingual graph
  of general knowledge,'' in \emph{Thirty-first AAAI conference on artificial
  intelligence}, 2017.

\bibitem{hsu2008combining}
M.-H. Hsu, M.-F. Tsai, and H.-H. Chen, ``Combining wordnet and conceptnet for
  automatic query expansion: A learning approach,'' in \emph{Information
  Retrieval Technology: 4th Asia Infomation Retrieval Symposium, AIRS 2008,
  Harbin, China, January 15-18, 2008 Revised Selected Papers 4}.\hskip 1em plus
  0.5em minus 0.4em\relax Springer, 2008, pp. 213--224.

\bibitem{fang2008re}
H.~Fang, ``A re-examination of query expansion using lexical resources,'' in
  \emph{proceedings of ACL-08: HLT}, 2008, pp. 139--147.

\bibitem{hsu2006query}
M.-H. Hsu, M.-F. Tsai, and H.-H. Chen, ``Query expansion with {ConceptNet} and
  {WordNet}: An intrinsic comparison,'' in \emph{Information Retrieval
  Technology: Third Asia Information Retrieval Symposium, AIRS 2006, Singapore,
  October 16-18, 2006. Proceedings 3}.\hskip 1em plus 0.5em minus 0.4em\relax
  Springer, 2006, pp. 1--13.

\bibitem{divya_elasticsearch}
M.~S. Divya and S.~K. Goyal, ``An advanced and quick search technique to handle
  voluminous data,'' \emph{Compusoft}, vol.~2, pp. 171--175, 2013.

\bibitem{zamfir_elasticsearch}
V.-A. Zamfir, M.~Carabas, C.~Carabas, and N.~Tapus, ``Systems monitoring and
  big data analysis using the elasticsearch system,'' in \emph{2019 22nd
  International Conference on Control Systems and Computer Science
  (CSCS)}.\hskip 1em plus 0.5em minus 0.4em\relax IEEE, 2019, pp. 188--193.

\bibitem{namboodiri2004online}
A.~M. Namboodiri and A.~K. Jain, ``Online handwritten script recognition,''
  \emph{IEEE Transactions on Pattern Analysis and Machine Intelligence},
  vol.~26, no.~1, pp. 124--130, 2004.

\bibitem{Doodle2App}
S.~Mohian and C.~Csallner, ``{Doodle2App}: Native app code by freehand {UI}
  sketching,'' in \emph{Proc. 7th IEEE/ACM International Conference on Mobile
  Software Engineering and Systems (MOBILESoft), Tool Demos and Mobile Apps
  Track}.\hskip 1em plus 0.5em minus 0.4em\relax ACM, May 2020, pp. 81--84.

\bibitem{sklearn_api}
L.~Buitinck, G.~Louppe, M.~Blondel, F.~Pedregosa, A.~Mueller, O.~Grisel,
  V.~Niculae, P.~Prettenhofer, A.~Gramfort, J.~Grobler, R.~Layton,
  J.~VanderPlas, A.~Joly, B.~Holt, and G.~Varoquaux, ``{API} design for machine
  learning software: experiences from the scikit-learn project,'' in \emph{ECML
  PKDD Workshop: Languages for Data Mining and Machine Learning}, 2013, pp.
  108--122.

\bibitem{huffman2007well}
S.~B. Huffman and M.~Hochster, ``How well does result relevance predict session
  satisfaction?'' in \emph{Proc. 30th Annual International {ACM} {SIGIR}
  Conference on Research and Development in Information Retrieval}.\hskip 1em
  plus 0.5em minus 0.4em\relax ACM, Jul. 2007, pp. 567--574.

\end{thebibliography}

\end{document}